\documentclass[11pt]{article}
\usepackage{authblk}
\usepackage{geometry}
\usepackage{amsmath,amsfonts,amssymb}
\usepackage{graphicx}
\usepackage{setspace}
\usepackage{xcolor}
\usepackage{tocloft}
\usepackage{tikz}
\usepackage{fp}
\usepackage{tikzscale}
\newcommand{\tm}[1]{\textrm{#1}}
\renewcommand{\d}{\, \text{d}}
\newcommand{\zb}[1]{\mbox{\boldmath{${#1}$}}}

\renewcommand{\d}{\, \text{d}}

\usepackage{fancyhdr}
\title{Moving Table Magnetic Particle Imaging: A stepwise approach preserving high spatio-temporal resolution}

\author[a,b,*]{Patryk Szwargulski}
\author[a,b]{Nadine Gdaniec}
\author[a,b]{Matthias Graeser}
\author[a,b]{Martin M\"oddel}
\author[a,b]{Florian Griese}
\author[c]{Kannan M. Krishnan}
\author[d]{Thorsten M. Buzug}
\author[a,b]{Tobias Knopp}
\affil[a]{Section for Biomedical Imaging, University Medical Center Hamburg-Eppendorf, 22529 Hamburg, Germany}
\affil[b]{Institute for Biomedical Imaging, Hamburg University of Technology, 21073 Hamburg, Germany}
\affil[c]{Materials Science and Engineering Department, University of Washington, Seattle, 98195 Washington, United States}
\affil[d]{Institute of Medical Engineering, University of L\"ubeck, 23562 L\"ubeck, Germany}
\affil[*]{Corresponding Author: p.szwargulski@uke.de }

\cftpagenumbersoff{figure}
\cftpagenumbersoff{table} 

\date{Received: 18 April 2018; Accepted: 1 November 2018; Published: 27 November 2018}
\begin{document} 
\maketitle

\begin{abstract}
Magnetic Particle Imaging (MPI) is a highly sensitive imaging method that enables the visualization of magnetic tracer materials with a temporal resolution of more than 46 volumes per second. In MPI the size of the field of view scales with the strengths of the applied magnetic fields. In clinical applications those strengths are limited by peripheral nerve stimulation, specific absorption rates, and the requirement to acquire images of high spatial resolution. Therefore, the size of the field of view is usually a few cubic centimeters. To bypass this limitation, additional focus fields and/or external object movements can be applied. In this work, the latter approach is investigated. An object is moved through the scanner bore one step at a time, while the MPI scanner continuously acquires data from its static field of view. Using a 3D phantom and dynamic 3D \textit{in vivo} data it is shown that the data from such a moving table experiment can be jointly reconstructed after reordering the data with respect to the stepwise object shifts and heart beat phases.
\end{abstract}


\section{Introduction}

Magnetic Particle Imaging (MPI) is a tomographic imaging modality that uses magnetic fields to image the spatial distribution of magnetic nanoparticles (MNPs) \cite{ref1,ref2}. The imaging method is based on the excitation of MNPs with one or several oscillating magnetic fields and the simultaneous detection of the MNP response using pick up coils. For spatial encoding a selection field is applied that can either have the form of a field-free point (FFP) or a field-free line (FFL). The excitation fields have a frequency in the kHz range and are limited in amplitude due to the risk of peripheral nerve stimulation (PNS) and/or specific absorption rate (SAR) \cite{ref3,ref4}. The gradient strength of the selection field is directly linked to the achievable spatial resolution \cite{ref5} such that it is usually chosen to be as high as possible. Since the size of the field of view (FoV) is determined by the ratio between the excitation field amplitude and the selection field gradient, the FoV is limited to few cubic centimeters in practice. One way to overcome the FoV limitation while preserving the spatial resolution is the measurement of multiple patches using additional focus fields \cite{ref6,ref7,ref8}, where the FoV is shifted to various positions and the individual imaging patches are combined either prior or post image reconstruction \cite{ref9,ref10}.
One major challenge for focus-field based multi-patch imaging sequences is that field imperfections lead to distortions when the FoV is shifted far away from the scanner center \cite{ref11}. 

One alternative to bypass the FoV limitation in MPI without the problems associated with field imperfections is to move the object through the FoV. Such methods are well known from established tomographic methods such as computed tomography (spiral CT) \cite{ref12} and magnetic resonance imaging (moving table MRI) \cite{ref13}. In MPI a moving table approach was used in several 1D MPI systems to increase the dimensionality to 2D or 3D \cite{ref14,ref15,ref16, ref17,ref18,ref35}. The aim of this work is to apply the linear moving table approach to 3D scans using Lissajous sampling trajectories. For that, the object is shifted through the 3D sampling volume while several Lissajous cycles are sampled at each table position \cite{ref19}. Reconstruction of such measurement data is non-trivial because the data from different Lissajous cycles have to be combined to a spatially and temporally consistent dataset prior to image reconstruction. This is necessary since the same part of the particle distribution may be present in several acquisition periods.

The Lissajous trajectories used in MPI are defined by orthogonal sinusoidal excitations with frequencies centered around 25~kHz as described in Ref.~\cite{ref1,ref36}. MPI is a fast imaging modality that allows to image about 46 volumes per second \cite{ref47} and thus, has a great potential for real-time imaging in small volumes. Nevertheless, a phase-specific reconstruction of e.g. \textit{in vivo} measurements is challenging due to cardiac and respiratory motion. For the standard MPI measurements already an adaption of the reconstruction is proposed to overcome this challenge \cite{ref31}. With enlarging FoV the temporal resolution gets lost, and only static or quasi static objects can be imaged without artifacts. Alternative scanner designs were able to image larger FoVs with a frame rate of 20 images per second \cite{ref46}. Until now, no methods were presented yet for enlarging the FoV while preserving the temporal resolution in MPI.

In this work, we propose a suitable reordering method to sort the raw data of one object passage through the imaging region into a multi-patch dataset \cite{ref19} that can be jointly reconstructed  \cite{ref7}.  Since the reordering method can be applied to static and to dynamic data, we perform two different experiments. 

First, we use a static 3D phantom to compare the multi-patch reconstruction of data acquired with the moving-table scan sequence with data acquired using the focus-field based sequence. This allows the comparison of the image quality using both methods and the analysis of the influence of the table movement on the proposed method. Further, we apply the moving-table scan sequence to a dynamic 3D \textit{in vivo} measurement of a mouse exhibiting a periodic heart beat. Using the reordering method we are able to reconstruct an image of the mouse in an enlarged FoV, while preserving the periodic heart beat motion.   

\section{Materials and Methods}

\subsection{Imaging Sequence}

\begin{figure}
\centering
a) \includegraphics[width=0.45\textwidth]{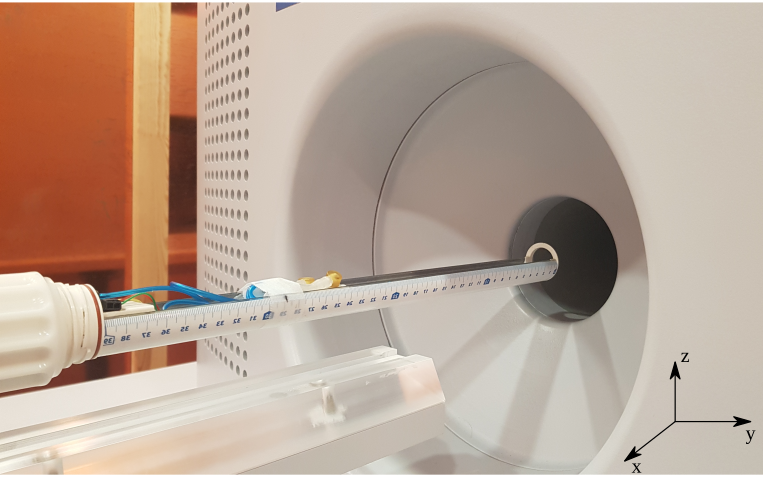}\quad b) \includegraphics[width=0.45\textwidth]{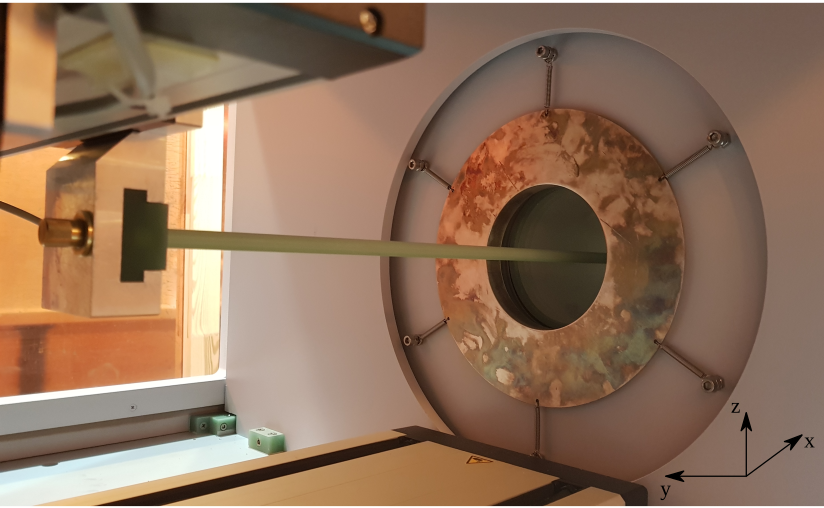}
\textcolor{white}{\rule{\textwidth}{0.5cm}}
c)\quad\includegraphics[width=0.9\textwidth]{Schema.pdf}
\caption{Visualization of the front a) and the back b) of the MPI-Scanner. In a) the animal bed used for the in vivo experiments is shown. In b) the robot rod used for the precise motion is visualized. In c) a schematic demonstration of the moving table experiment is shown. Thereby the mouse lies on the table and is moved through the imaging center inside of the scanner.}
\label{fig:Schema}
\end{figure}

We consider an imaging sequence with continuous acquisition of a 3D Lissajous trajectory in the center of the scanner. Thereby one full cycle of the trajectory is defined as one raw data frame. In this work, we use a moving table approach to enlarge the FoV of a single 3D measurement along the bore and not to increase the dimensionality of the image from 1D or 2D to 3D. The volumetric imaging sequence allows us to measure overlapping areas along the bore while keeping the isotropic spatial resolution. The spatial overlap also leads to a temporal oversampling of the structures enabling an analysis of periodic motions of the scanned object. The object is mounted on a table that can be moved through the scanner, shown in Fig.\,\ref{fig:Schema}. In our case the table is an animal bed from MINERVE (Equipment Veterinaire Minerve) shown in Fig.\,\ref{fig:Schema} a) and a linear three axis robot from ISEL Fig.\,\ref{fig:Schema} b). Within this work, we will drop frames acquired during table movement. To yield a high duty cycle, the table is moved stepwise through the imaging center, as schematically shown in Fig.\,\ref{fig:Schema} c). In this way, only those frames experiencing a movement need to be dropped while the frames where the table rests can be used for image reconstruction. The length of each raw-data frame in our case is 21.5~ms while the table movement takes approximately $>$100\,ms. To reach a high duty cycle, the resting phase at each table position is chosen longer than the movement phase such that at each position several continuous frames are available. In case of static objects this allows an improvement of the signal quality using block-wise averaging, whereas in case of a dynamic object the frames can be used to capture the full dynamic of the object.   

\subsection{Signal Equation} \label{Sec:SignalEq}
Next, we will investigate how the MPI signal model has to be adapted to take the moving table into account.
The MPI signal equation is based on electromagnetic induction, where one or several receive coils pick up the magnetization changes of the MNP \cite{ref1,ref5}. For a single receive coil the induced voltage $u(t)$ is given by
\begin{equation} \label{Eq:staticImagingEq}
	u(t)=-\mu_0\int_{\mathbb R^3} \frac{\partial}{\partial t} \zb{M}(\zb{r},t) \cdot \zb{p}(\zb{r})\d^3r,
\end{equation}
where $\mu_0$ denotes the vacuum permeability, $\zb{p}(\zb{r})$ denotes the receive coil sensitivity and \linebreak[4] ${\zb{M}(\zb{r},t)=c(\zb{r})\zb{m}(\zb{r},t)}$ is the net magnetization caused by a certain distribution $c(\zb r)$ of effective magnetic moments $\zb{m}(\zb{r},t)$. 

In case of a moving table experiment a total number of $L$ periods were measured. The object position within period $l = 1, \dots, L$ is given by $\zb{x}_l$ such that the particle distribution $c$ is sampled at shifted positions $(\zb r - \zb x_l)$. Here, a static object is assumed during the excitation period. Frames with table movement will be dropped in a post-processing step.
The signal equation in period $l$ is thus given by
\begin{align}\label{Eq:DynImaging}
	u_l(t) &= -\mu_0 \int_{\mathbb R^3}  \frac{\partial}{\partial t}\zb{m}(\zb{r},t) \cdot \zb{p}(\zb{r}) c(\zb{r}-\zb{x}_l)\d^3r\notag \\
   & =\int_{\mathbb R^3}  s(\zb{r},t) c(\zb{r}-\zb{x}_l)\d^3r
\end{align}
where $s(\zb r,t) = (-\mu_0)\frac{\partial}{\partial t}\zb{m}(\zb{r},t) \cdot \zb{p}(\zb{r})$ denotes the MPI system function. Since the time signal $u_l(t)$ is usually not directly accessible in MPI it is common to expand the periodic signal into a Fourier series with coefficients $\hat{u}_{l,k}$ \cite{ref30, ref29, ref28, ref39}. The MPI signal will be available at most frequencies $f_k$, whereas the coefficients of the excitation frequencies have to be dropped. Due to finite sampling of the time signal, the frequency index is limited to $K$ such that $k$ runs from $1$ to $K$. $u_l(t)$ does not contain a DC offset so that the index $k=0$ can be neglected. After Fourier expansion Eq. \eqref{Eq:DynImaging} can be written as
\begin{equation} \label{Eq:DynImagingEqFD}
\hat{u}_{l,k}=\int_{\mathbb R^3} \hat{s}_{k}(\zb{r}) c(\zb{r}-\zb{x}_l) \d^3r,
\end{equation}
where $\hat{s}_{k}(\zb{r})$ is the Fourier expansion of $s(\zb r,t)$ with respect to the time variable $t$.
By applying the substitution $\tilde{\zb r} = \zb{r}-\zb{x}_l $ Eq.\,\eqref{Eq:DynImagingEqFD} can be rewritten as
\begin{equation} \label{Eq:DynImagingEqFDFinal}
\hat{u}_{l,k}=\int_{\mathbb R^3} \hat{s}_{k}(\tilde{\zb r}+\zb{x}_l) c(\tilde{\zb r}) \d^3\tilde{r}.
\end{equation}
This integral equation has the form of a linear integral operator that takes as input the continuous particle distribution $c$ and maps it to the frequency components $\hat{u}_{l,k}$. 

\subsection{Data Processing}\label{subsec:DataProc}
\begin{figure}
\centering
\includegraphics[width=0.7\textwidth]{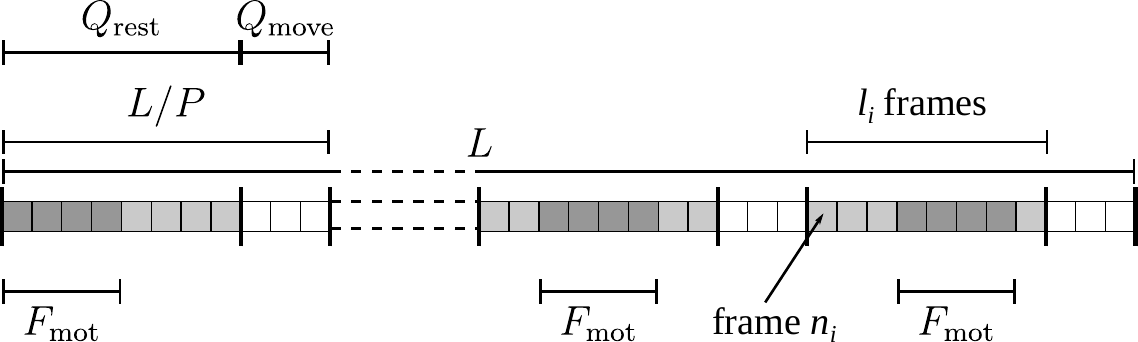}
\caption{Data processing. The data stream of length $L$ is divided into $P$ sections of length $Q_{\mathrm{move}}+Q_{\mathrm{rest}}$. During the acquisition of the raw-data frames $Q_{\mathrm{rest}}$, the object is resting at one spatial position. Each group contains $l_i$ raw-data frames with $n_i$ as the first frame of each group. The object is moved to the next position during the acquisition of the raw-data frames $Q_{\mathrm{move}}$. The raw-data frames $F_{\mathrm{mot}}$ are part of $Q_{\mathrm{rest}}$ and are used for the reconstruction in case of periodic motion of the object.}
\label{fig:timeseries}
\end{figure}

To process data obtained from a moving table experiment one has to take the varying object locations $\zb x_l$ into account. The data stream with a total number of $L$ raw-data frames is divided into $P$ groups of length  
\begin{equation}
Q_\text{move} + Q_\text{rest} = \frac{L}{P},
\end{equation}
where for simplicity we assume that $P$ is a divider of $L$. The table is at rest at a single position during the acquisition of the $Q_{\mathrm{rest}}$ frames, while the table is moving to the next position during the acquisition of the $Q_{\mathrm{move}}$ frames. This is illustrated in Fig. \ref{fig:timeseries}. 
We now group the raw-data frames by their table position. The $i$-th group with $i=1,\dots,P$ contains raw-data frames $$l_i=n_i,\dots,n_i + Q_\text{rest}-1,$$ where $$n_i:=(i-1)(Q_\text{move} + Q_\text{rest}) + 1.$$ 

The frames with table movement are dropped prior to reconstruction to prevent motion artifacts. When considering constant object shifts $\zb \xi$, the positions can be expressed as \linebreak[4] $\zb x_{l_i}=(i-1)\zb{\xi} + \zb{\xi}_\text{start}$, where $\zb{\xi}_\text{start}$ is the initial position. The impact of the step size on the reconstruction result will be studied in section \ref{Sec:Results}.

\begin{figure}
\centering
\includegraphics[width=1\textwidth]{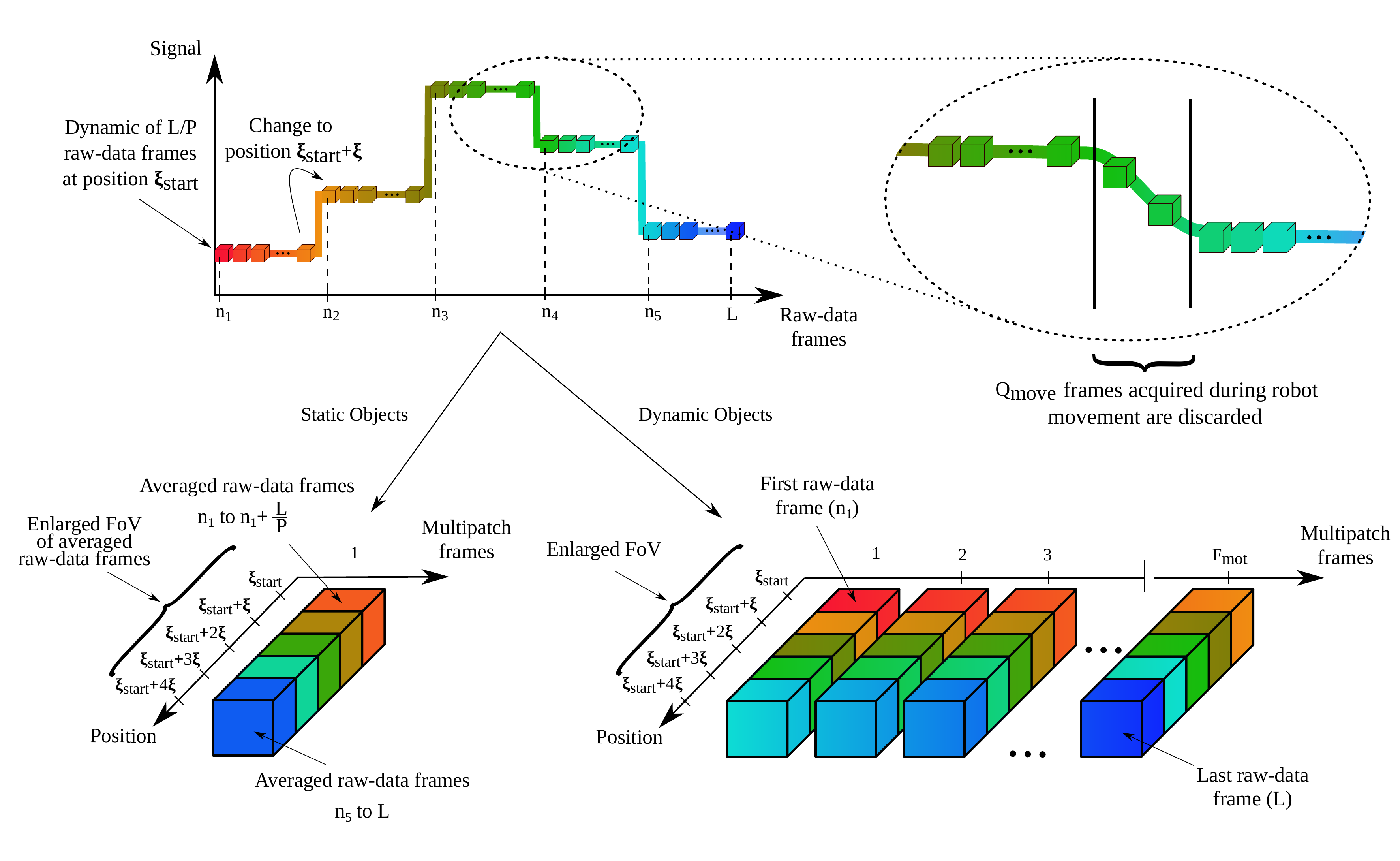}
\caption{The signal is continuously acquired while the object was moved to $P=5$ different positions. The plot on the top left sketches the signal of a single frequency component throughout the $L$ measurements and the grouping into $L/P$ consecutive raw-data frames. The color codes the time point at which the measurement was obtained from red (first) to blue (last). On top right the movement of the robot between two groups indicating the raw-data frames $Q_\textrm{move}$ is shown. At the bottom the data processing for the static and the dynamic case are shown. Thereby on the left the averaging and on the right the resorted data are shown, where the correspondence between each multi-patch frame and the position is given. }
\label{fig:Skizze}
\end{figure}

In the following, we will first consider the data processing for static objects. Next, we will explain how to reorder the data of dynamic 3D objects to a spatially and temporally consistent dataset. A dataset that is spatially and temporally consistent can be reconstructed frame by frame on an enlarged FoV.

\subsubsection{Static Objects}
To generate consistent data of the measured static objects one raw-data frame of each group P can be selected and combined into a multi-patch dataset. However, for static objects all frames $Q_\text{rest}$ at each position represent the same state. In order to improve the signal to noise ratio (SNR) of the measurement signal one can apply block averaging
\begin{equation}
\tilde{u}_{i,k}=\frac{1}{Q_\text{rest}}\sum^{n_i+Q_\text{rest}-1}_{l=n_i}\hat{u}_{l,k},
\end{equation}
where $\tilde{u}_{i,k}$ is the block averaged signal of group $i$ for $i=1,\dots,P$. The block averaging case is graphically sketched in Fig. \ref{fig:Skizze} at the bottom left. The resulting dataset includes the averaged data of each group corresponding to a single frame and will be referred in following as a multi-patch frame.

\subsubsection{Objects with Periodic Motion}
\label{subsec:DynamicObjects}
For objects with periodic motion, the proposed block averaging leads to a coherent dataset, but the dynamic information gets lost. To preserve those information one has to generate a dynamic 3D dataset consisting of $F_\text{mot}$ multi-patch frames. Thereby it is assumed that the motion of the object is periodic, such that for each group at table location $i=1,...,P$ a raw-data frame $m_{i,j}$ with $j=1,...,F_\text{mot}$ exists. For that, the number of measured frames in one motion period $F_\text{mot}$ has to be smaller than $F_\text{mot} \leq Q_{\mathrm{rest}}/2$ to ensure that the information of the full dynamic is included in each group, compare Fig.~\ref{fig:timeseries}. Since the motion period does not match the length of the raw-data frames of each group ($Q_\textrm{rest}$) the raw-data frames corresponding to the same state of motion are shifted for each location. The shift for the frame group $i$ and the motion state $j$ is given by 
\begin{equation}
m_{i,j}:=\lceil n_i/F_{\mathrm{mot}}\rceil\,F_{\mathrm{mot}}+j-1,
\label{Eq:MotionState}
\end{equation}
with the ceiling operator defined by $\lceil \iota \rceil:=\min\{\tau \in \mathbb{Z}\,|\tau \geq \iota\}$. For the reconstruction of one full motion cycle all $F_\text{mot}$ frames are used. If the resting time $Q_\text{rest}$ is long enough to contain $\alpha -1$ full motion cycles $ \alpha F_\text{mot} \leq Q_\text{rest}, \alpha \in \mathbb{Z}$, and $\alpha \geq 3$ the additional data can be used to increase the duration of the time series or can be averaged to reconstruct a single period of the motion series with an improved SNR. The scan efficiency can be determined by $1-Q_\text{lost}/Q_\text{rest}$ with the unused raw-data frames $Q_\text{lost}$ given by
\begin{equation}
Q_\text{lost}=Q_\text{rest}-\left(\left \lceil \frac{Q_\text{rest}-(m_{i,1}-1)}{F_\text{mot}}\right \rceil-1\right) F_\text{mot}.
\end{equation}
The scan efficiency thus ranges from 1/2 to 1.
The corresponding signal is given by 
\begin{equation}
   \tilde{u}_{i,k}^j = \hat{u}_{m_{i,j},k}.
\end{equation}
Each of the $j=1,\dots,F_{\mathrm{mot}}$ motion states can be reconstructed individually yielding a dynamic 3D volume with $F_{\mathrm{mot}}$ multi-patch frames. The dynamic case is graphically sketched in Fig. \ref{fig:Skizze} at the bottom right.

\subsection{Image Reconstruction}

The coordinate transform applied in section \ref{Sec:SignalEq} has brought the moving table image equation into the same form as the data from a multi-patch experiment. Instead of the object, the system function is shifted in space in multi-patch experiments. Thus, for image reconstruction one can use the joint multi-patch reconstruction algorithm proposed in Ref.\,\cite{ref7}. Discretization of \eqref{Eq:DynImagingEqFDFinal} at $N$ equidistant positions $\zb r_n$, $n=1,\dots,N$ laying on a 3D grid yields a linear system of equations
\begin{equation} \label{Eq:MVEq}
\begin{pmatrix}
\zb{S}_1\\
\zb{S}_2\\
\vdots\\
\zb{S}_{P}
\end{pmatrix}
\zb c=
\begin{pmatrix}
\zb{\hat{u}}_1\\
\zb{\hat{u}}_2\\
\vdots\\
\zb{\hat{u}}_{P}
\end{pmatrix},
\end{equation}
where $\zb c = (c(\zb r_n))_{n=1}^{N}$ is the particle concentration vector, $$\zb{S}_i = ( \hat{s}_{k}(\tilde{\zb r}+\zb{x}_{l_i}) )_{k=1,\dots,K;n=1,\dots,N}$$
is the shifted system matrix, and $\zb{\hat{u}}_i = (\tilde{u}_{i,k})_{k=1}^{K}$ is either the averaged measurement vector or one  measurement vector $\zb{\hat{u}}_i^j = (\tilde{u}_{i,k}^j)_{k=1}^{K}$ of the dynamic 3D dataset. 

The linear system \eqref{Eq:MVEq} can be solved by iterative solvers such as the Kaczmarz algorithm, which is applied in this work \cite{ref26}. The Kaczmarz algorithm sweeps over the rows of the system matrix and adds a certain fraction of each matrix row to the temporary solution vector, which is iteratively updated.  

The system matrices $\zb{S}_i$ can be derived from a central system matrix $\zb{S}_{\text{center}}$ by shifting the individual matrix row with respect to the spatial shift $\zb{x}_{l_i}$. This shift can be applied on-the-fly during the iteration process such that only a single system matrix $\zb{S}_{\text{center}}$ has to be kept in memory. 

The central system matrix $\zb{S}_{\text{center}}$ can be obtained by a calibration procedure shifting a delta sample through the region of interest while continuously measuring the signal response in the receive coils. The volume chosen for this calibration procedure -- named system function field-of-view (SF-FoV) -- is usually chosen larger than the drive-field field-of-view (DF-FoV) since the support of the system function is slightly larger than the DF-FoV \cite{ref20}. The DF-FoV is the region that is captured by the trajectory of the FFP during one Lissajous cycle. Especially for joint multi-patch reconstruction it is crucial to measure the system function with sufficient support along the stitching direction of the patches since otherwise truncation artifacts would occur \cite{ref8}.

\subsection{Experimental Setup}

Experiments were carried out using a pre-clinical FFP MPI scanner (Bruker Biospin MRI GmbH Ettlingen, Germany) \cite{ref21}. Instead of the build-in receive coils a custom gradiometer-based coil was used improving the SNR of the measured signal~\cite{ref22}. Measurements were performed using excitation field amplitudes of $12\,$mT/$\mu_0$ in all three directions. The selection field gradient was set to \linebreak[4] $\zb G=\text{diag}(-1,-1,2)\,\tm{T/m/$\mu_0$}$ resulting in a DF-FoV of size $24\times 24\times 12\,\tm{mm}^3$.

The system matrices were measured on a grid of $35\,\times\,25\,\times\,15$ positions with a delta sample of size $2\,\times\,2\,\times\,1\,\tm{mm}^3$. For the phantom experiments the tracer Perimag (micromod Partikeltechnologie GmbH, Germany) with a concentration of 151\,mmol(Fe)/L was used. The system matrix for the \textit{in vivo} experiment was measured with the tracer LS-008 \cite{ref23,ref24} at a concentration of 92.03\,mmol(Fe)/L.

Two different experiments were performed to validate the proposed moving table acquisition sequence. First, a static object is measured to investigate how many table positions are required for artifact-free reconstruction. Next, the static object is used to compare the proposed method to the focus field based approach. Finally, dynamic \textit{in vivo} experiments were performed to demonstrate the frame resorting algorithm.

\subsection{Phantom Experiments}
\begin{figure}
\centering
a)\includegraphics[width=0.6\textwidth]{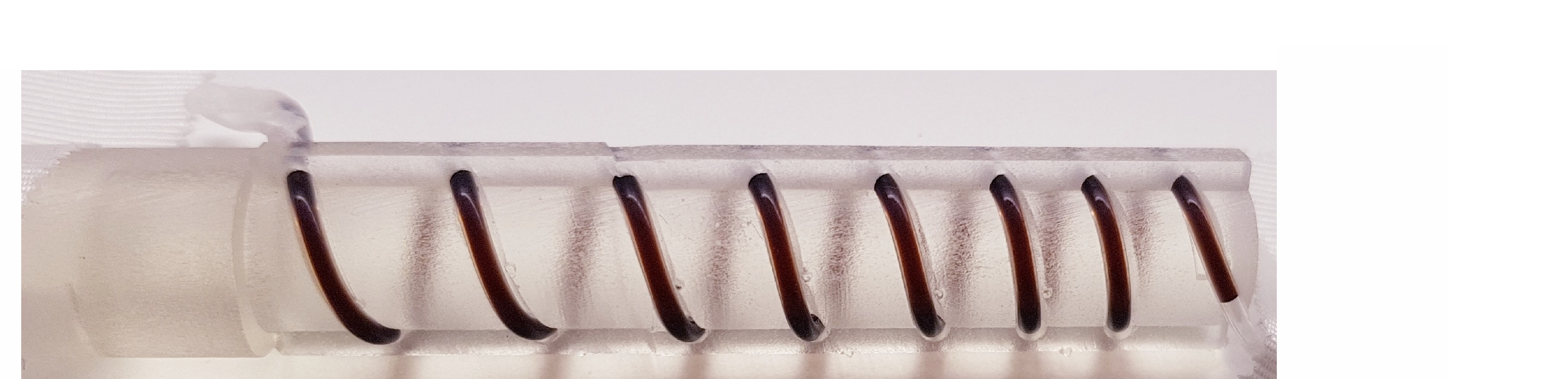}\\
b)\includegraphics[width=0.6\textwidth]{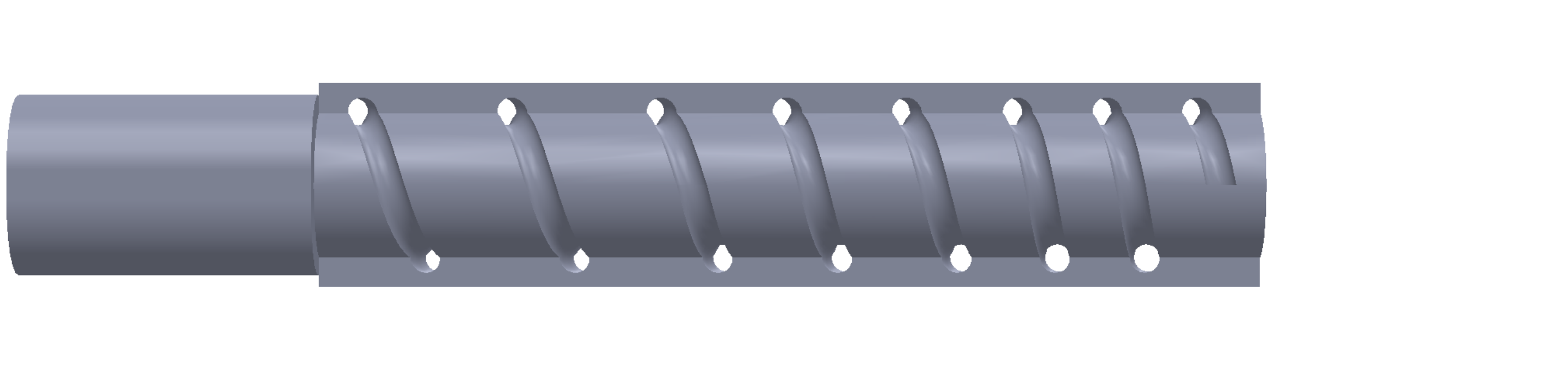}\\
c)\includegraphics[width=0.6\textwidth]{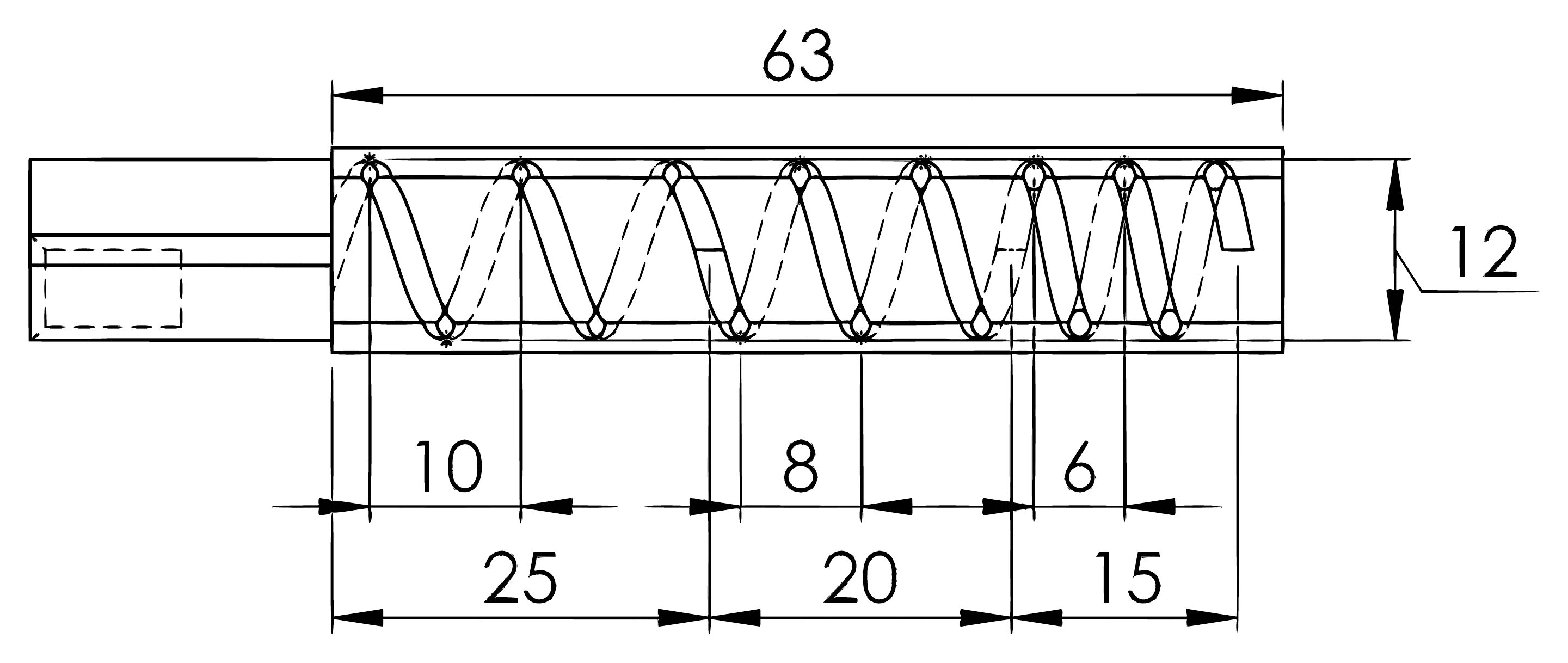}
\caption{Visualization of the spiral phantom: a) shows a photo of the phantom while b) shows a CAD model, and c) shows a technical drawing including the dimensions in cm. }
\label{fig:SpiraleModel}
\end{figure}

For the static phantom, a flexible tube was wound around a 3D-printed holder. The tube had an inner diameter of $0.813\,$mm and was filled to a length of about $28\,$cm with Perimag (micromod Partikeltechnologie GmbH, Germany). The tracer solution was diluted to a concentration of $75.9\,$mmol(Fe)/L. A CAD model and a picture of the phantom are shown in Fig.~\ref{fig:SpiraleModel}. The phantom was moved with a robot to 65 positions with a step size of 1\,mm covering a total FoV of $90\times 24\times 12\,\tm{mm}^3$. Since the length of the DF-FoV is $24\,\tm{mm}$ in $x$-direction, the measured volumes are highly overlapping. In order to study how much overlap is necessary for successful reconstruction, additional measurements were performed. The step size was successively increased from 2\,mm to 40\,mm in 2\,mm steps. During the measurements $Q_\text{rest}$ was kept constant at about 110 frames corresponding to 2.4\,s duration. The number of frames used in the reconstruction was adapted to ensure constant signal averaging for each step size.   

The number of frames between two steps, $Q_\text{move}$, depends on the spatial distance between to steps and varies between 11 and 40 frames. This corresponds to 0.24\,s and 0.86\,s, since a single frame has a repetition time of 21.54\,ms. The scan times range from  
5.6\,s for the measurement with a 40\,mm step size up to about 3 minutes for the measurement with a step size of 1\,mm, robot movement included.      

For comparison, measurements using focus fields were performed. For the focus field measurements, the center of the phantom was placed in the middle of the bore and the full range of focus field amplitude in $x$-direction was applied. Thirty-five focus fields were used, ranging from -17\,mT to 17\,mT in 1 mT increments. With the applied selection field gradient $\zb G$ each increment corresponds to a shift of 1 mm, resulting in a $58 \times 24 \times 12 \textrm{mm}^3$ FoV. At each position 100 repetitions were performed and the data were averaged prior to the reconstruction.

\subsection{\textit{In vivo} Experiments}

The \textit{in vivo} experiment was performed using a healthy mouse approved by local animal care committees (Beh\"orde f\"ur Lebensmittelsicherheit and Veterin\"arwesen Hamburg, Nr. 42/14,70/14,16/41). A similar workflow as outlined in our previous publication Ref.\,\cite{ref25} was applied. The mouse was anesthetized using isoflurane and put on the mouse bed mounted at a manual positioning system that allows movement of the mouse-bed through the scanner bore. Fiducials were used to adjust the height of the mouse bed such that the mouse passes the DF-FoV.

After mouse placement $10\,\mu\tm{L}$ of the long circulating tracer LS-008 \cite{ref23,ref24} with a concentration of $92.03\,$mmol(Fe)/L was injected. Then the MPI measurement was started and the mouse was moved to six positions with a step size of 1.1\,cm covering a FoV of $90\,\times\, 24\,\times\,12\,\tm{mm}^3$. The step-size of 1.1\,cm was selected to achieve a suitable overscan of the data avoiding gaps between the patches. Since the mouse has to be positioned manually, the movements were performed by hand using a ruler with a precision of 1\,mm. Due to the missing correspondence between frame number and muse bed position this correspondence was restored in a post processing step. The raw data stream was analyzed using Fourier methods at a certain frequency component (102.6\,kHz). Furthermore, we analyzed the reconstructed single patch images frame by frame. Both enabled to determine the frames $n_i$ at which the resting phase started. The mean number of frames $Q_\text{rest}$ was 2648 corresponding approximately to 57\,s. The mean time for the manual shift $Q_\text{move}$ was about 45 frames corresponding to 1\,s. The number of frames $F_\text{mot}$ containing the full dynamic of the cardiac motion was determined to be 8 frames. The total scan time was about 5\,min and 48\,s, manual shift included. In this work we reconstructed 80 frames from the entire dataset to show the principle of our method. 

\subsection{Reconstruction Parameters}

For the reconstruction a regularized form of the Kaczmarz algorithm \cite{ref26} was used with a preselected number of frequencies based on the SNR of the system matrix \cite{ref26,ref27,ref28,ref29,ref30}. The reconstruction is parameterized by three parameters: the SNR threshold for frequency selection, the number of iterations, and the Tikhonov regularization parameter. The latter is usually given relative to the normalized Frobenius norm of the system matrix \cite{ref30}. All three parameters were manually selected based on visual inspection of the contrast of the reconstructed images. Automatic selection of MPI reconstruction parameters is still an open problem, since the reconstruction has several parameters (SNR threshold, Tikhonov regularization parameter, and the number of iterations) that influence each other and in turn cannot be optimized independently. Nevertheless, a iteration number of 3 and a relative regularization parameter of 0.01 was kept constant for all reconstructions. Only the SNR threshold was different for the experiments since the tracer concentration was lower for the \textit{in vivo} experiments. In case of the phantom experiments an SNR threshold of 3 was chosen corresponding to 3157 selected frequency components. In case of the \textit{in vivo} experiment the SNR threshold was set to 40 corresponding to 674 selected frequency components. 

Since in the moving table approach the position of the DF-FOV is constant, only one system matrix is required for reconstruction. For the reconstruction of the focus field data we reused the same system matrix to reduce the calibration effort. We note that explicitly measured system matrices at each focus field position might improve the reconstruction results.

The implementation of the reconstruction is realized in the programming language julia \cite{ref40,ref41} using self-developed software and open source packages \cite{ref42,ref43}. All reconstructions are performed on a workstation equipped with two Intel(R) Xeon(R) CPU E5-2640 v3 CPUs running at 2.6\,GHz and a main memory of 512\,GB. The reconstruction time for the phantom experiments was 35\,s in the worst case of 1\,mm step size and 43\,s reconstructing 80 multi-patch frames for the \textit{in vivo} data. 

\section{Results} \label{Sec:Results}

\subsection{Phantom Experiments}

\begin{figure}
\centering
\includegraphics[width=0.8\textwidth]{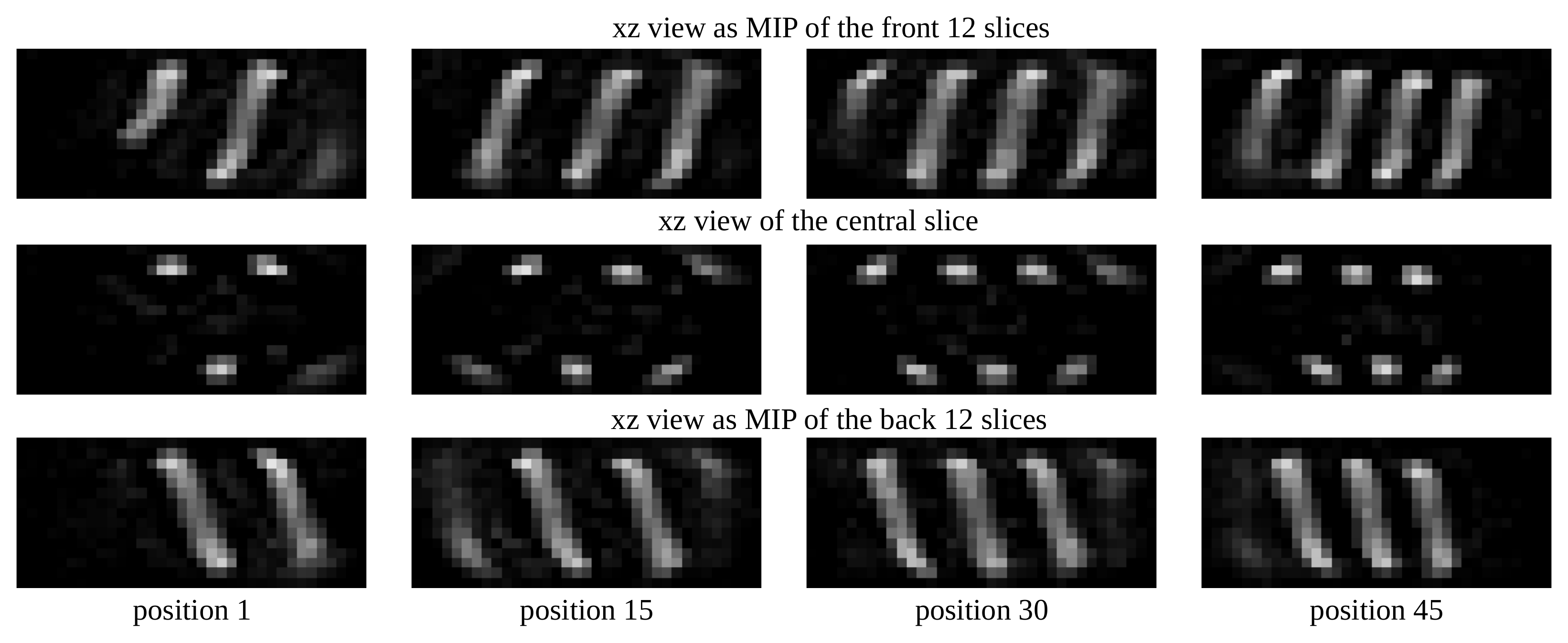}
\caption{Reconstruction results of the spiral phantom reconstructed for several table position separately. For visualization 4 equidistantly selected positions are shown. In the first and the last row a maximum-intensity projection of the twelve front and back $y$-slices are shown, respectively. In the middle row only the central slice through the spiral is visualized.
}
\label{fig:SpiraleErgSingle}
\end{figure}

Reconstruction results of the data reconstructed for several positions are shown in Fig.~\ref{fig:SpiraleErgSingle}. The upper and lower row contain maximum-intensity projections (MIP) along the $y$-direction of the first twelve and last twelve slices of the reconstruction results, respectively. The middle row shows the central slice of the reconstructed images. The images from one column correspond to the same spatial shift along the $x$-direction. From the 65 reconstructed datasets images at four selected positions are shown. One can see that each reconstructed dataset only captures a fraction of the entire spiral. At the left and right edges of the images the signal drops since those areas are outside the DF-FOV.

\begin{figure}
\centering
\includegraphics[width=0.99\textwidth]{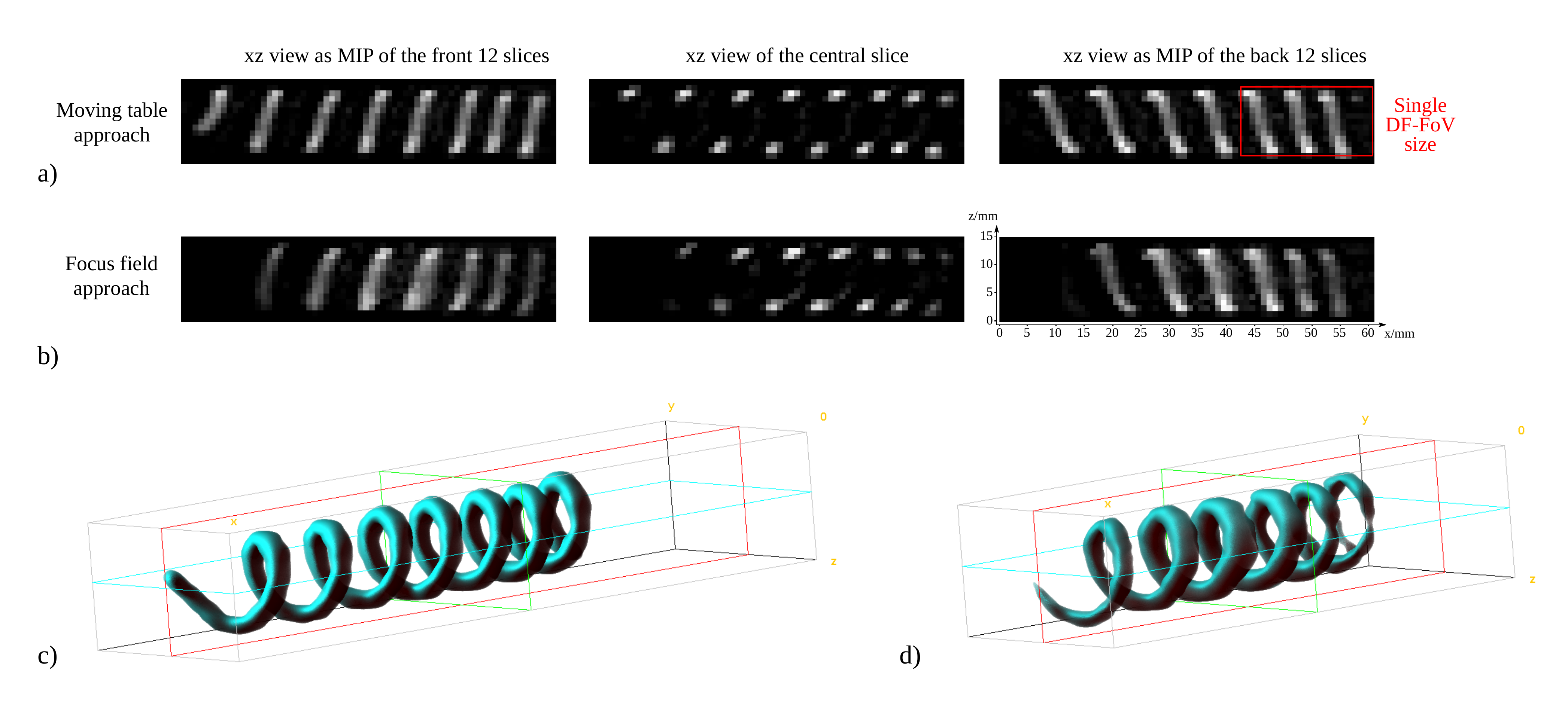}
\caption{Reconstruction results of the static spiral phantom are shown in a)-d). In a) the reconstruction results of the data acquired using the proposed moving table approach. In b) the result of the reconstruction of the focus field data is shown. In both cases on top and bottom a maximum intensity projection of the twelve front and back $y$-slices are shown. In the center of a) and b) the central slide is shown. Using the moving table approach the FoV could be extended to 90\,mm in the $x$ direction while the extension was limited to 58\,mm for the focus field approach. For better comparison the reconstruction results of the moving table approach are cropped to the size of the focus field reconstructions.  To visualize enlargement of the FoV through the scanner bore, the size of the DF-FoV is shown for reference in a). Volume renderings of the spiral phantom are shown, reconstructed using the moving table approach c) and the focus field approach d). For the volume rendering the data were thresholded with 15\,\%  of the maximum pixel value and cubically interpolated. }
\label{fig:SpiraleErg}
\end{figure}

Reconstruction results of the joint multi-patch reconstruction for the phantom dataset using the data from all 65 shift positions are shown in Fig.~\ref{fig:SpiraleErg}. In (a) the reconstruction results of the data acquired using the moving table approach are given and in (b) the reconstruction results of the data acquired using the focus field-based approach are shown. In the first and last column the maximum-intensity projections (MIP) along the $y$-direction of the first twelve and last twelve slices of the reconstruction results are presented. The middle row shows the central slice of the reconstructed images. A volume rendering of the reconstruction results are given in (c) and (d). The volume rendering was performed on thresholded data using a cubic interpolation. One can clearly see that the spiral is successfully reconstructed using the moving table approach. No artifacts at the patch boundaries are visible since the joint reconstruction enforces a smooth and consistent reconstruction. The reconstruction of the focus field data seem to be less homogeneous at the edges and more than one full winding of the spiral is missing. This is directly linked to the limitation of the applicable focus field amplitude since the DF-FoV could not be shifted wide enough to scan the whole phantom.

\begin{figure}
\centering
\includegraphics[width=0.99\textwidth]{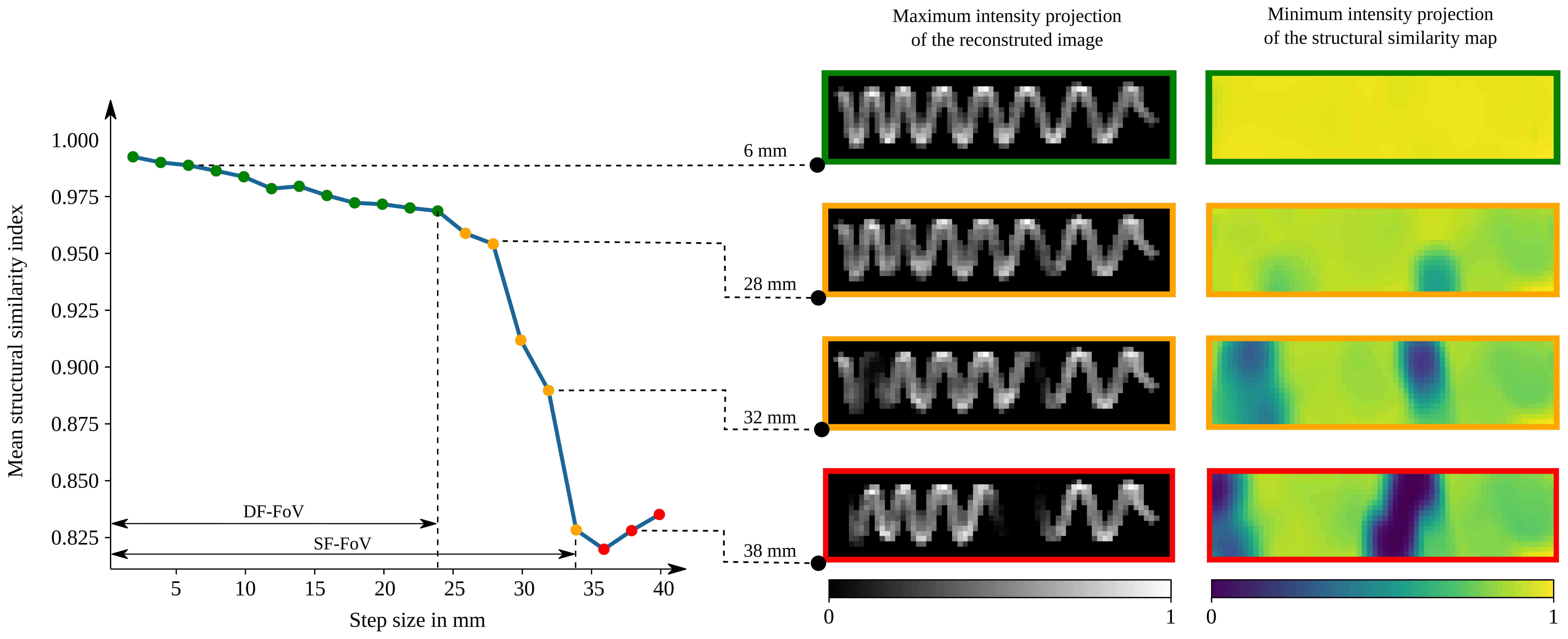}
\caption{Influence of the step size on the image quality. The mean structural similarity index (SSIM) between the reconstruction results for different step sizes and the reconstruction result with a step size of $1\,$mm is shown on the left. Reconstructed images with different step sizes are given on the right. Additional the corresponding SSIM map for the chosen reconstructions is shown at the right border. The step sizes can be divided into three different groups. Overlapping DF-FoVs of neighboring patches are indicated in green, while overlapping SF-FoVs are orange. The cases indicated in red have neither overlapping DF-FoVs, nor SF-FoVs.}
\label{fig:SpiraleStepsErg}
\end{figure}

The reconstruction results of the step size analysis are shown in Fig.~\ref{fig:SpiraleStepsErg}. The step size was increased from 2~mm until 40~mm. The range of step sizes can be divided into three different groups. In the first group (indicated in green) the DF-FoVs of neighboring patches overlap or are stitched at the boundaries. In the second group (indicated in orange) the SF-FoV of neighboring patches overlap while there is a gap between the DF-FoVs. In the last group (indicated in red) neither the DF-FoV nor the SF-FoV do overlap. As would be expected, the spiral can fully be reconstructed if the step size is less then or equal to the DF-FoV size. Additionally, even with small gaps between the DF-FoVs images could be reconstructed with visually negligible error. If the step size is larger, the reconstruction results of the spiral exhibit gaps at the patch boundaries. To quantify these results the structural similarly index (SSIM) \cite{ref37,ref38,ref45} between the normalized reconstructed images was calculated. The SSIM calculates a map of local deviations between two images ranging from -1 (different) to 1 (identical)\cite{ref45}. The mean of this map can be used to quantify the similarity. The reconstructed image using a step size of 1\,mm and all averaged frames shown in Fig.~\ref{fig:SpiraleErg} was used as the ground truth. To ensure a constant averaging for the remaining step sizes, the number of used raw-data frames over all positions was kept constant.

The results of the analysis are shown in Fig.~\ref{fig:SpiraleStepsErg}. The mean SSIM for different step sizes is given in the left part of Fig.~\ref{fig:SpiraleStepsErg}. One can see that in case of overlapping DF-FoVs the deviation remains below 5\,\%. Even for the step sizes of 26\,mm and 28\,mm corresponding to a gap between the DF-FoVs the numerical error remains below 5\,\%. The similarity decreases rapidly down to 83\,\% for a step size over 28\,mm. The reconstruction results (middle part) and the minimum intensity projections of the SSIM maps (right part) give a visual impression of the numerical results of the mean SSIM. The SSIM maps visualize the spatial deviation from the ground truth. The SSIM map of the 6\,mm step size is nearly homogeneous, whereas the similarity decreases with the step size. Especially, for the reconstruction results acquired with a step size larger than 28\,mm some gaps in the spiral are visible. For the data acquired with a step size of 28\,mm the deviation is already slightly visible in the SSIM map, but there is no visible gap in the spiral. The deviations of the last three reconstruction results (step size of 36\,mm to 40\,mm) are all around 83\,\%, while the mean similarity slightly increases. In this cases only two patches were acquired with increasing gap size in between. This can be explained by the fact that the SSIM, as a structural measure, takes the density of the spiral into account. Since the density of the turns increases along the spiral the similarity is better matched for the dense part compared to the sparse part. 

\subsection{\textit{In vivo} Experiments}
\begin{figure}[htb!]
\centering
\includegraphics[width=0.9\textwidth]{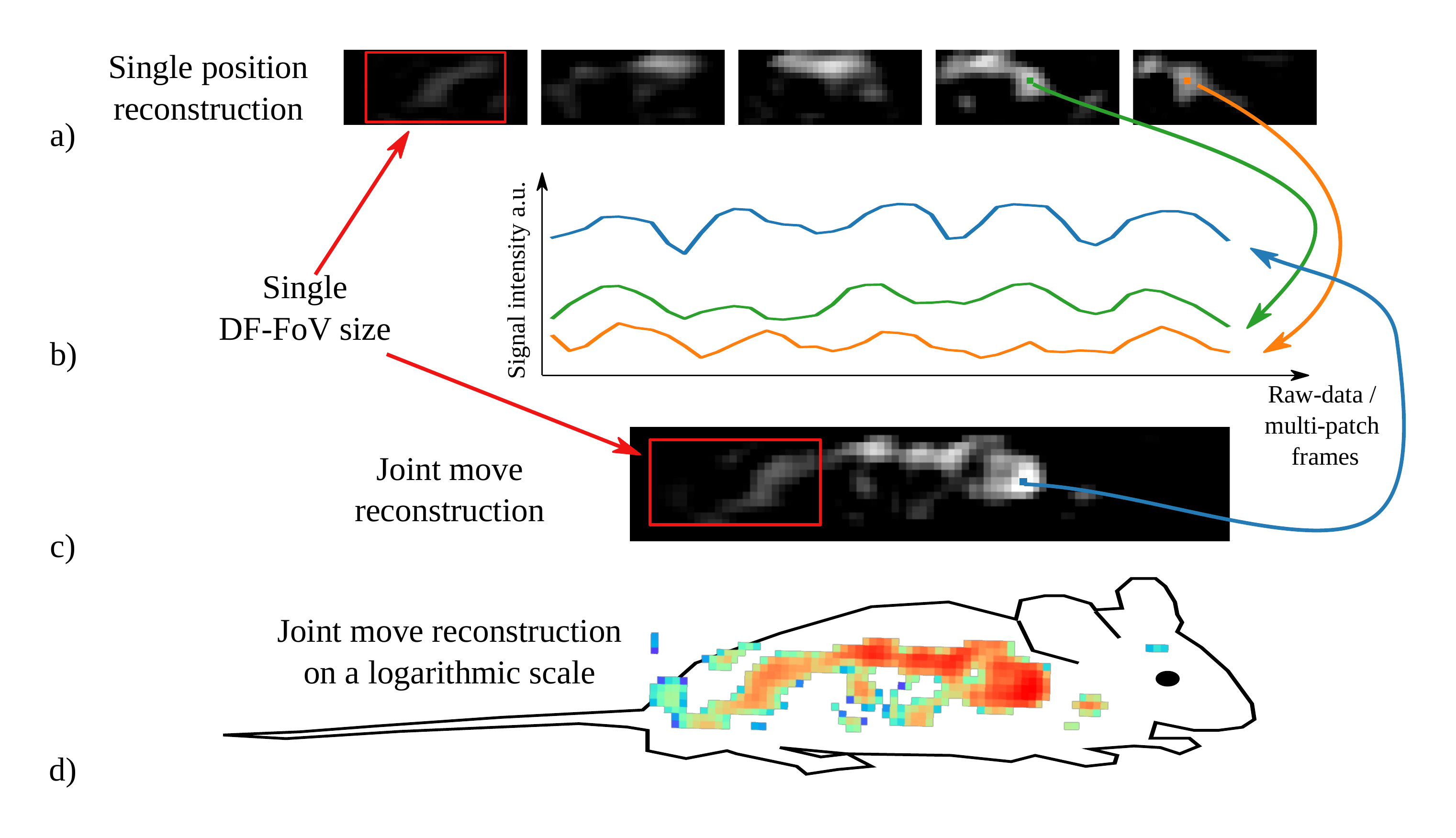}
\caption{Reconstruction results of the \textit{in vivo} experiment. Separate reconstruction results are given in (a). The reconstruction result of the joint multi-patch reconstruction is given in (c). In (b) the time-intensity curve for the selected pixels are given. One can see that the phase of the single and the joint reconstruction match. For better visualization the joint reconstructed image is shown on a logarithmic scale in (d) and is schematically overlaid on a mouse model.}
\label{fig:MausLangsam}
\end{figure}

\begin{figure}[htb!]
\centering
\includegraphics[width=0.9\textwidth]{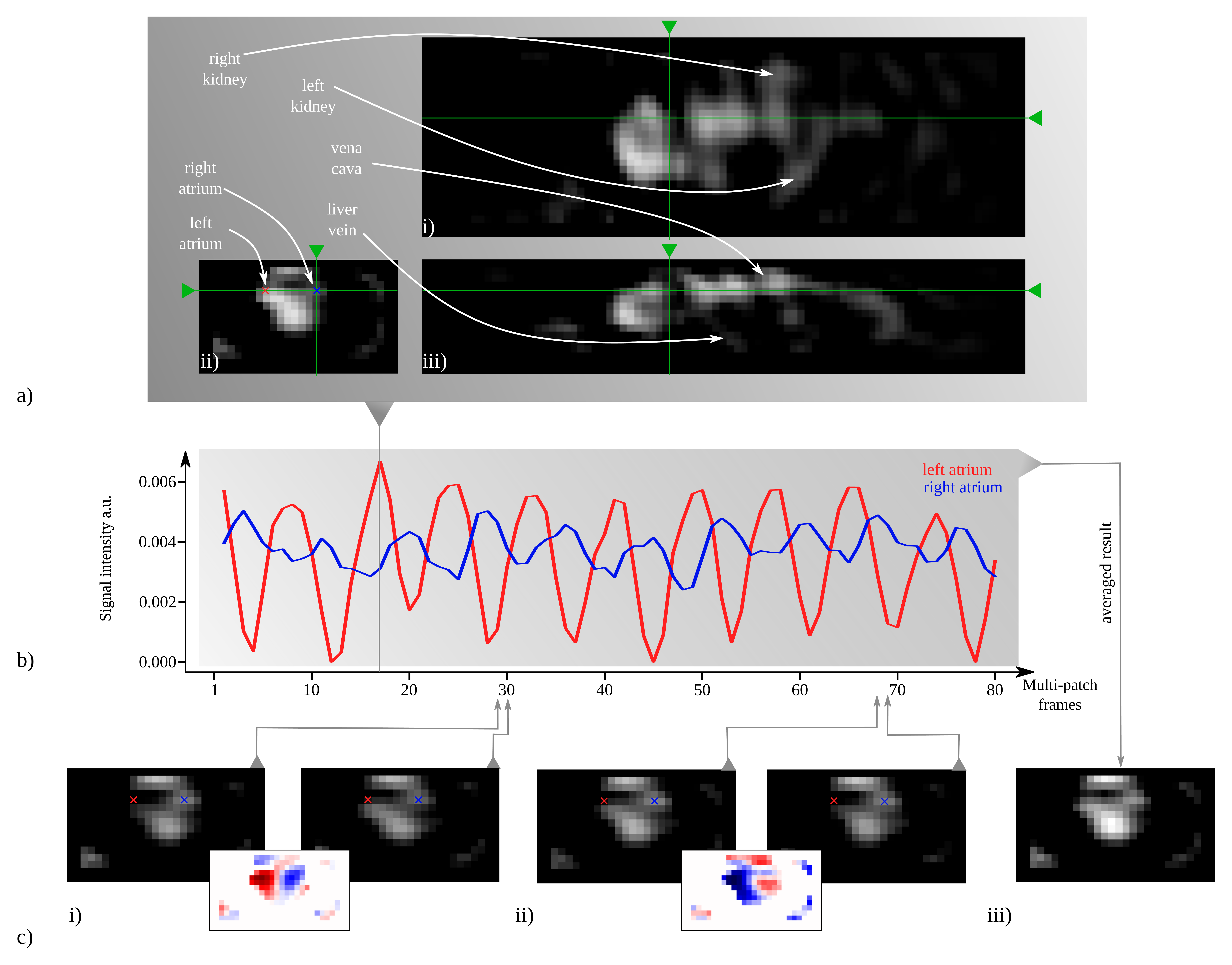}
\caption{In a) the views of the coronal (i), transverse (ii), and sagittal (iii) plane of the 17th frame from the dynamic 3D reconstruction result of the mouse are shown. In the images multiple anatomical structures are marked. The position of the slides in the different planes are pictured in green. In b) the graph of the signal intensity over time of the left and right atrium are shown. In c) the transverse plane of the 29th and 30th frame (i) and the 69th and 70th (ii) frame are visualized. In between, the differences of both images are presented. Each image shows the heart of the mouse. Across the frames it is visible how the heart pumps the MNP. In (c) (iii) the transverse plane of the averaged data of all 80 frames is shown. (Supplemental Video available). }
\label{fig:MausPointed}
\end{figure}

Reconstruction results of the \textit{in vivo} experiment are shown in Fig. \ref{fig:MausLangsam} and Fig. \ref{fig:MausPointed}. For the reconstruction results given in Fig.~\ref{fig:MausLangsam}~(a) the dynamic 3D data were reconstructed separately for each spatial shift of the entire object. The central slice of the reconstructed image is shown, crossing the vena cava, the heart, the aorta, and the brain.
Data processing as described in section \ref{subsec:DynamicObjects} was applied to the dynamic \textit{in vivo} data to ensure that the signal from different spatial shifts are in phase of the heart beat. For verification, the time-intensity curve for the marked pixels is given in (b) representing the same spatial location inside the body, but different global shifts of the entire object.  
The reconstruction result of the joint multi-patch reconstruction is given in (c) together with an exemplary DF-FoV marked in red. Since the MPI data cover a dynamic range of several magnitudes, the reconstruction result is also shown on a logarithmic scale (d) for better visual inspection using a more fine-grained colormap.

In Fig.\,\ref{fig:MausPointed} (a) the coronal (i), transverse (ii), and sagittal (iii) plane of the reconstruction results are shown. The images represent the 17th frame of the reconstructed 80 frames. In the different planes the kidneys, the liver vein, and the vena cava are visible. In the sagittal plane (i) at the selected position the vena cava is visible continuously from heart to tail. Furthermore, in Fig.\,\ref{fig:MausPointed} (b) the temporal progression of the intensity of the reconstruction result at the positions of the left and right atrium is given. In the graph the phase shift effected by the heart beat is shown. In (c) the 29th and 30th multi-patch frame (i) and in (ii) the 69th and 70th multi-patch frame of the transverse plane are shown representing the different phases of the heart beat. A difference image between the frames is included to emphasize the changes between them. The last part of Fig.\ref{fig:MausPointed} (c) (iii), depicts the reconstruction of the average over all 80 frames.

\section{Discussion}
In this work, we enlarged the FoV in MPI more than three times in one direction using a moving table approach while there is theoretically no limitation. The acquisition time was less than 2 seconds, though table shifts and the acquisition of multiple measurements per table position were not accounted for. The shift time is strongly dependent on the step size and the hardware used to perform the movement. In our phantom experiments using a step size of 1\,mm and the ISEL robot for movement the mean shift time was about 236.9\,ms corresponding to 11 frames. Measuring 110 frames at each position the duty cycle was about 90\,\%. In summary, this leads to a total scan time of 2\,min and 52\,s. The duty cycle for the \textit{in vivo} experiment was much higher, since long resting times of about 57\,s was chosen between two steps to get a high amount of data. This supports the conclusion that the movement time is not an issue for a step wise moving table experiment.  

A moving table approach is commonly used for x-space approaches. In terms of spatial and temporal resolution, and FoV extend a comparison is given in the following. The reported scan time in Ref.~\cite{ref44} for a 3D volume scan using the x-space approach was reported to be 570\,s. Thereby a FFP scanner was used and a FoV of $20\times 41\times 100$\,mm$^3$ was scanned. In Ref.~\cite{ref44} an imaging sequence and reconstruction process was proposed that leads to a comparable isotropic spatial resolution to our method. If we would extend our measurement procedure to cover the same region in all three dimensions, it would take approximately 47.4\,s. For the calculation of this time we used the worst case time for the robot movement, and an averaging of 100 raw-data frames to increase the SNR of the raw data signal.

In this work, measurements of a static 3D phantom and dynamic \textit{in vivo} measurements were performed. In both cases the step size was chosen smaller than the size of the DF-FoV ensuring overlap between patches. Further, the influence of the step size between 2\,mm and 40\,mm on the image quality was investigated. Based on the reconstructed images it could be shown that the results are compatible for step sizes up to the size of the DF-FoV. Further, even small gaps between the patches could be reconstructed with visually negligible errors. Nevertheless, to avoid gaps in the reconstructed images the step size should be equal or smaller to the size of the DF-FoV.

The moving table approach was compared to the focus field approach on static 3D data. While both methods were able to reconstruct the center of the phantom correctly, slight distortions appear with raising focus field amplitudes. In both cases, the reconstructions were performed using a single system matrix. Explicitly measured system matrices at each patch position might improve the reconstruction result, but this would result in a huge time expenditure. Nevertheless, in the current system the focus field amplitude is technically limited to $\pm 17$\,mT. With the applied gradient strength the extent of the FoV is limited to 58\,mm and can not capture the whole phantom leading to about one missing turn of the spiral. The main advantage of the proposed method is the theoretically unlimited enlargement of the FoV along the scanner bore. 

The \textit{in vivo} measurements showed that the moving table method can be used in the presence of a superimposed periodic motion such as a heart beat by synchronizing the motion phases for different table positions. Although we corrected only the heart beat the method is not restricted to cardiac motion, but in principal also applicable to respiratory motion. The synchronization of the table position, the measured frames, and the motion period is a necessary condition, otherwise motion artifacts will occur due to inconsistencies in the data. In our work, we did correct for the phase shifts at a scale of the length of a singular measurement cycle. Note that this method can be further refined by using subframe data handling methods as proposed in Ref.~\cite{ref31}. Apart from synchronization, adopted data averaging techniques as proposed in Ref.~\cite{ref31} can be applied to increase the SNR of the temporally resolved data. Further, it has to be kept in mind that the table movement during the \textit{in vivo} experiment was performed manually and the data were not synchronized with the measurement. This could be improved by developing an adapter to use the external robot for movement of the animal bed and synchronization of the MPI measurements with the animals cardiac and respiratory cycles. Especially for applications demanding a high precision in positioning this will be a crucial factor.

The effective enlargement of the FoV was achieved by reordering the data into multi-patch datasets in a post processing step, which does not allow for real-time reconstruction of the measured data. The latter could be achieved by reordering the data during acquisition. Furthermore, the data acquired during object movement was not used for image reconstruction. The efficiency of moving-table MPI could be improved by incorporating these data into the reconstruction. To this end, the motion needs to be taken into account for in the reconstruction to avoid motion artifacts \cite{ref32}. One possible solution could be the application of temporally dependent shifts to the time domain representation of the system function \cite{ref48}.

In summary, the focus field approach and the moving table approach can be used to increase the FoV in MPI. Both change the spatial location of the object relative to the location of the drive-field. Thus, the same reconstruction techniques can be applied. While the workflow in the focus field approach is easier because the object does not need to be moved, the hardware requirements are higher. Focus-field MPI scanners can either be made by simple coil configurations that require huge electrical power losses if the FFP is shifted far off-center \cite{ref33}, or there exist smarter coil setups as developed in Ref.\,\cite{ref33,ref34}. However, these require a larger number of coils and power amplifiers. These setups will suffer from large field non-linearities requiring correction techniques to reduce artifacts in the reconstructed image. The moving table approach thus might be a simple alternative to bypass the challenges of focus-field MPI scanners. 

The moving table and the focus-field approach are not mutually exclusive. It would also be possible to combine both approaches in a single device and scanning protocol. For example in a clinical scenario approximately $10\,\times\,10\,\times\,10$ patches are required to capture the human heart with the scanning parameters used in this work. The increase of the FoV in direction of the scanner bore could be realized by moving the table and the increase in the perpendicular plane by shifting the DF-FoV with the help of focus fields. This would considerably reduce the hardware demands compared to a 3D focus-field scanner since the focus-field channel along the bore can be replaced by the table movement still allowing for full 3D imaging in a large FoV.

\section{Conclusion}

In conclusion, the proposed method offers an alternative for enlarging the FoV in MPI to the usage of focus fields. It enables to reconstruct temporally resolved data in an enlarged FoV without artifacts at the patch boundaries and potentially unlimited extend in direction of the scanner bore. In phantom and \textit{in vivo} experiments it was shown that the moving table approach makes it possible to increase the FoV and still preserve the temporal resolution in case of objects experiencing periodic motion.

\subsection*{Disclosures}
The authors state no conflict of interest and have nothing to disclose.

\subsection*{Acknowledgments}
P.S., N.G., M.G., M.M., F.G. and T.K. thankfully acknowledge the financial support by the German Research Foundation (DFG, grant number KN 1108/2-1) and the Federal Ministry of Education and Research (BMBF, grant number 05M16GKA). T.M.B. thankfully acknowledge the financial support by the German Research Foundation (DFG, grant number BU 1436/9-1). We thank P. Ludewig for animal handling during the animal experiments. We also thank the anonymous reviewer for their very helpful comments that helped us improving the manuscript.

\bibliography{ms}   

\vspace{1ex}
\noindent Biographies and photographs of the other authors are not available.

\end{document}